\documentclass[english,prd,a4paper,onecolumn,nofootinbib]{revtex4}
\pdfoutput=1
\usepackage{graphicx, bm, color, babel}
\usepackage{hyperref}
\usepackage{amsmath}
\usepackage{color}

\def\be{\begin{equation}}
\def\ee{\end{equation}}
\def\bea{\begin{eqnarray}}
\def\eea{\end{eqnarray}}

\newcommand{\<}{\langle}
\renewcommand{\>}{\rangle}

\newcommand{\ga}{\gamma}
\newcommand{\De}{\Delta}
\newcommand{\de}{\delta}
\newcommand{\ep}{\epsilon}
\newcommand{\ka}{\kappa}
\newcommand{\Om}{\Omega}

\newcommand{\bn}{{\bm{n}}}

\newcommand{\bx}{\mathbf{ x}}

\newcommand{\avd}[1]{\left\<\,\overline{#1} \,\right\>}

\newcommand{\no}{\mathbf{n}_{o}}
\newcommand{\xo}{\mathbf{x}_{o}}

\newcommand\spart{\;\raise1.0pt\hbox{$/$}\hskip-6pt\partial}
\newcommand\spartb{\;\overline{\raise1.0pt\hbox{$/$}\hskip-6pt
\partial}}

\newcommand{\DD}{{\cal D}}
\newcommand{\FF}{{\cal F}}
\newcommand{\OO}{{\cal O}}

\newcommand{\bfe}{\mathbf{e}}
\newcommand{\bal}{\boldsymbol{\alpha}}
\newcommand{\bnabla}{\boldsymbol{\nabla}}

\newcommand{\kap}{\langle \kappa_1^2\rangle}

\begin{document}

\title{\vspace{-0.5cm}{\normalsize \normalfont \flushright CERN-PH-TH-2015-075\\}
\vspace{0.6cm}
Cosmological ensemble  and directional averages of observables 
}
\author{Camille Bonvin$^1$}
\author{Chris Clarkson$^2$}
\author{Ruth Durrer$^3$}
\author{Roy Maartens$^{4,5}$}
\author{Obinna Umeh$^4$}
\affiliation{$^1$Theory Division, CERN, 1211 Geneva, Switzerland\\
$^2$Astrophysics, Cosmology \& Gravity Centre and Department of Mathematcis \& Applied Mathematics, University of Cape Town, Cape Town 7701, South Africa\\
$^3$D\'epartement de Physique Th\'eorique \& Center for Astroparticle Physics, Universit\'e de Gen\`eve, Quai E.\ Ansermet 24, CH-1211 Gen\`eve 4, Switzerland\\
$^4$Physics Department, University of the Western Cape, Cape Town 7535, South Africa\\
$^5$ Institute of Cosmology \& Gravitation, University of Portsmouth, Portsmouth PO1 3FX, United Kingdom\\
{\rm E-mail: camille.bonvin@cern.ch, chris.clarkson@gmail.com, ruth.durrer@unige.ch, roy.maartens@gmail.com, umeobinna@gmail.com}
} \vspace*{0.2cm}

\date{\today}

\begin{abstract} 
We show that at second order, ensemble averages of observables and directional averages do not commute due to gravitational lensing -- observing the same thing in many directions over the sky is not the same as taking an ensemble average. In principle this non-commutativity is significant for a variety of quantities that we often use as observables and can lead to a bias in parameter estimation. We derive the relation between the ensemble average and the directional average of an observable, at second order in perturbation theory. We discuss the relevance of these two types of averages for making predictions of cosmological observables, focusing on observables related to distances and magnitudes. In particular, we show that the ensemble average of the distance in a given observed direction is increased by gravitational lensing, whereas the directional average of the distance is decreased. For a generic observable, there exists a particular function of the observable that is not affected by second-order lensing perturbations. We also show that standard areas have an advantage over standard rulers, and we discuss the subtleties involved in averaging in the case of supernova observations.
 
\end{abstract}

\maketitle

\section{Introduction}

Cosmological observations have become very precise. Especially for the analysis of the cosmic microwave background (CMB) data one has to take into account not only first-order perturbations but also second-order effects like lensing~\cite{Lewis:2006fu,cmblensing, cmb,cmb2nd}.  For other perturbed quantities like supernova distances and redshifts as a function of observed 
direction~\cite{BenDayan:2012ct,chris1,BenDayan:2012wi,chris,giov,Ben-Dayan:2014swa}, cosmic shear~\cite{shear,bisp} and galaxy number counts~\cite{Bertacca:2014dra,Bertacca:2014wga,Yoo:2014sfa,DiDio:2014lka}, second-order perturbative expressions have recently been published and demonstrated to be possibly non-negligible. We need to include these second-order effects if we want to compare theory with very precise observations. Their measurement is also an opportunity to test general relativity since most of these effects are different in theories which modify gravity.

In this paper we show that special attention has to be given when comparing a second-order calculation to observations. In cosmology, since we have only one universe at our disposition, we often replace ensemble averages by averages over directions. Here we show that at second order, directional and ensemble average do not commute. This means that the ergodic assumption is broken by observation on the observer's past light-cone: due to gravitational lensing, observing the same thing in many directions over the sky is not the same as taking an ensemble average. 

The existence of different kinds of averages has already been discussed by Kibble and Lieu~\cite{kibble}, for the particular case of the magnification $\mu$. They argued that the average over random directions in the sky is not the same as the average over a random distribution of sources. They showed that the `random-source average' of the magnification $\mu$ is exactly given by its background value -- a result previously demonstrated by Weinberg~\cite{weinberg} -- but that 
the `random-direction average' of the magnification is affected by perturbations. They found that the quantity that is invariant under random-direction average is the reciprocal magnification $\mu^{-1}$.  

Here we extend this distinction to arbitrary observables. We argue that theoretically, we can only calculate ensemble averages, whereas observationally we usually average over directions. To compare theory with observation, we need therefore to first take the directional average of the observable and then the ensemble average. We give an explicit expression for the difference between this procedure and the ensemble average in a single line of sight. Our calculation is valid up to second order in perturbation theory and in the regime of weak lensing\footnote{We do not discuss here the more difficult problem of how caustics may affect observables~\cite{ellis}.}.

We apply our formalism to various observables. In particular we discuss the case of the distance, showing that the directional average of the distance (followed by an ensemble average) is {\it smaller} than its background value, whereas the ensemble average of the distance along a single line of sight is {\it larger} than its background value. We also discuss the case of isolated standard candles and standard rulers for which it is not so clear which averaging procedure to consider, since we do not usually have many sources at a given redshift. We illustrate the different biases in different variables and we discuss how to construct observables which minimise the bias from second-order perturbations. Finally we apply our formalism to the CMB angular power spectrum. We show that in multipole-space, the average over directions is automatically taken before the ensemble average. 

The rest of the paper is organised as follows: in section~\ref{s:gen} we show that ensemble average and directional average do not commute. We derive a general relation between the two types of averages, valid at second order in perturbation theory. In section~\ref{s:ex}, we discuss various examples and in section~\ref{s:con} we conclude.
 
\section{Averaging at second order}\label{s:gen}

We consider a cosmological experiment, i.e. an observation where we detect photons coming in at direction $\bn_o$ from a source situated at redshift $z$. Our observable can be the density of galaxies in a certain direction, the temperature of the CMB, the luminosity of a supernova, etc. We usually repeat the measurement over various directions in the sky, and we measure the mean of the observable and/or the correlation functions (averaged over all directions at a fixed angular separation).

To compare these measurements with theoretical predictions we have to perform ensemble averages. Cosmological perturbations are stochastic fields and only their ensemble average and their variance (or other higher-order correlation functions) can be calculated. The usual procedure is to assume that due to stochastic isotropy and the ergodic principle, the (measured) directional average of our observable is equal to its ensemble average. 
Here we show that this procedure is correct only at first order in perturbation theory. At second order, ensemble averages and averaging over directions are two distinct procedures, which do not commute.

\begin{figure}[!t]
\centering
\includegraphics[width=0.15\textwidth]{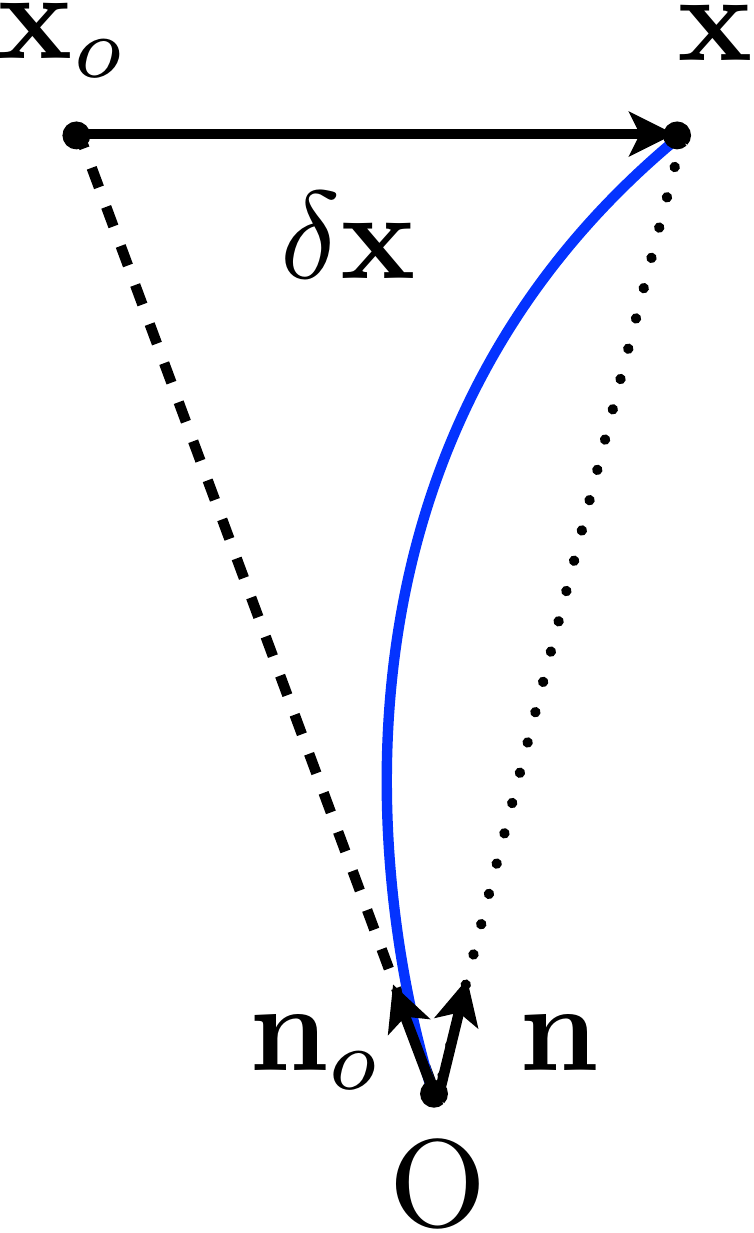}
\caption{\label{fig:positions} We measure the value of the observable $f$ at the true comoving position of the source $\bx$. This position is observed by photons coming in at the direction $\no$, corresponding to an unperturbed position in the sky $\xo\equiv\chi\,\no$. The deflection vector $\delta\bx$ relates $\xo$ to $\bx$ and the deflection angle is defined through $\bal=\delta\bx/\chi$.}
\end{figure}

We consider an arbitrary function of direction $f(\bn_o)$ in the sky. Here $\bn_o$ is the (lensed) observed direction. In an unperturbed universe, $f$ does not depend on directions: $f=f_0$. In a perturbed universe, we expand $f$ around $f_0$, in perturbations of order $\ep$ 
\be
\label{fno}
f(\bn_o) = f_0\left[1 +\ep \de_1(\bn_o) + \frac{\ep^2}{2}\de_2(\bn_o) +{\cal O}(3)\right] \,.
\ee
Taking the expectation value of~\eqref{fno} and assuming Gaussianity (so that the expectation value of third-order perturbations vanishes) we get
\be
\langle f(\bn_o) \rangle = f_0\left[1 +\ep \langle \de_1(\bn_o)\rangle + \frac{\ep^2}{2}\langle \de_2(\bn_o)\rangle +{\cal O}(4)\right] \,.
\ee
Naively we would set $\langle \de_1(\bn_o)\rangle=0$. However, at second order in perturbation theory this is not correct. 
When we go to second order, we have to take into account that the observed photon direction $\bn_o$ is lensed from the unperturbed source direction (see Fig.~\ref{fig:positions}),  $\bn=\bn_o+\bal$, where $\bal=\ep\bal_1 + \ep^2\bal_2/2 +{\cal O}(3)$ is the  deflection angle, which we assume to be small. As a consequence, the distribution of  images is not statistically homogeneous and isotropic, meaning that $\langle \de_1(\bn_o)\rangle\neq 0$. More precisely we have
\be
\ep\langle \de_1(\bn_o)\rangle=\ep\langle\de_1(\bn-\bal)\rangle= \ep\langle\de_1(\bn) \rangle-\ep^2\left\langle\bal_1\cdot\bnabla \de_1(\bn)\right\rangle+ {\cal O}(3)= -\ep^2\left\langle\bal_1\cdot\bnabla\de_1(\bn_o)\right\rangle+{\cal O}(3)\,.
\ee
In the last equality, we have used the fact that the distribution of the sources is statistically homogeneous and isotropic (a consequence of the statistical homogeneity and isotropy of the primordial fluctuations), so that $\langle \de_1(\bn)\rangle= 0$. Note that in expressions which are already second order, we do not need to distinguish between $\bn$ and $\bn_o$. 

The expectation value of the second-order expression for $f$ is therefore of the form
\be\label{e:ff1}
\< f(\bn_o)\> = f_0\left[1 -\ep^2\left\<\bal_1\cdot\bnabla \de_1(\bn_o)\right\> + \frac{\ep^2}{2}\< \de_2\> +{\cal O}(4)\right] \,.
\ee
The second term in~\eqref{e:ff1} directly follows from the non-random distribution of the images generated by gravitational lensing. This term can be simplified using
\be
\label{correction}
\left\<\bal_1\cdot\bnabla\de_1(\bn_o) \right\>=\left\<\bnabla\cdot\big(\bal_1 \de_1\big)( \bn_o) \right\>-\left\<\bnabla\cdot\bal_1(\bn_o) \de_1\right\>
=2\left\< \kappa_1 \delta_1\right\>\, ,
\ee
where the second equality follows since a total divergence does not contribute to the average and since the first-order convergence is given by $-2\ka_1 = \bnabla\cdot\bal_1(\bn_o)$ (see~\cite{cmb2nd} for details). With this, the expectation value of $f$ becomes
\be
\label{fav}
\< f(\bn_o)\>=f_0 \left[1-2\ep^2\left\<\ka_1 \de_1 \right\> + \frac{\ep^2}{2}\< \de_2\> +{\cal O}(4)\right]\, .
\ee
Let us now see what happens if instead of calculating the expectation value of $f$, we first average over observed directions, and then take the ensemble average. From~\eqref{fno} we have\footnote{Here we assume for simplicity that we average over the whole sky, but the argument holds for any patch of the sky.
The boundary terms which appear are vectors which disappear after ensemble averaging.}
\be
\label{avdf}
\avd{f}\equiv\frac{1}{4\pi} \left\<\int d\Om_{\bn_o}f (\bn_o)\right\>= f_0\left[1 +\ep \avd{\de_1(\bn_o)} + \frac{\ep^2}{2}\avd{\de_2(\bn_o)} +{\cal O}(3)\right] \, .
\ee

The first-order term gives
\bea
\avd{\de_1(\bn_o)}&=& \frac{1}{4\pi}\left\<\int d\Omega_{\bn_o} \de_1(\bn_o)\right\>
=\frac{1}{4\pi}\left\<\int d\Omega_{\bn} \left|\frac{\partial{\bn_o}}{\partial\bn} \right|\de_1(\bn-\bal)\right\>\nonumber\\
&\simeq& \frac{1}{4\pi}\left\<\int d\Omega_{\bn} \Big[ \de_1(\bn)-\epsilon\de_1(\bn)\bnabla\cdot\bal_1(\bn)-\epsilon\bal_1\cdot\bnabla \de_1(\bn)\Big]\right\>\, . \label{d1av}
\eea
The second and third terms can be combined into a total derivative. Since $\bn$ is unperturbed, the directional average over $\bn$ commutes with the ensemble average, and we obtain
\be
\avd{\de_1(\bn_o)}\simeq \frac{1}{4\pi}\int d\Omega_{\bn} \Big[ \left\<\de_1(\bn)\right\>-\epsilon\bnabla\cdot \left\<\de_1(\bn)\bal_1\right\>\Big]=0\, .
\ee
This result holds also at next order in perturbation theory. It relies only on the fact that the expectation value of total derivatives vanishes in a statistically homogeneous and isotropic universe\footnote{Note that the result does not hold if we enter the strong lensing regime where caustics and multiple images appear, and where the map $\bn_o \mapsto \bn$ is not one to one. However, the regions where this happens contribute a negligible area for most purposes.}.

For the second-order term in~\eqref{avdf}, we may neglect the perturbations of the direction (which contribute at third order only) so that $\de_2(\bn_o)\simeq\de_2(\bn)$, and directional and ensemble average of $\de_2$ commute
\be
\avd{\de_2(\bn_o)}=\langle \de_2\rangle  +{\cal O}(4) \, .
\ee
With this we obtain
\be\label{e:fbar}
\avd{f}=  f_0\left[1 +\frac{\ep^2}{2}\<\de_2 \>  +{\cal O}(4)\right] = \<f(\bn)\>\,.
\ee 
Comparing~\eqref{e:fbar} with~\eqref{fav} we see that
\be\label{e:fcomp}
 \<f(\bn)\>\ =\avd{f} = \< f(\bn_o)\>+ 2f_0\ep^2\left\<\ka_1\de_1 \right\> +{\cal O}(4)\,.
\ee
This equation provides a general relation between the ensemble and the directional average of any observable, valid at second order in perturbation theory, in the weak lensing regime.  In particular, it shows that ensemble and directional averages commute only if the variable under consideration has vanishing correlation with the convergence (or similarly with the deflection angle). An important consequence is that an observable whose ensemble average is invariant under second-order perturbations (like for example the magnification $\mu$, see~\cite{kibble,weinberg}) will automatically receive corrections when we take its directional average (and vice-versa).

In practice, if we measure a directional average, $\overline f$, the ensemble average which we have to compare our measurements with  is $\avd{f}$.  However, if we measure just one realisation $f(\bn_o)$ in a fixed observed direction $\bn_o$, we should compare our measurements with 
$\<f (\bn_o)\>$. 

Before we discuss important examples, let us also note the following: if we observe a power of $f$, say $f^p$, a Taylor expansion up to second order yields
\bea
\< f^p (\bn)\>=\avd{ f^p} &=& f_0^p\left[1 +\frac{p\ep^2}{2}\<\de_2 \>  +\frac{p(p-1)\ep^2}{2}\< \de_1^2\>  +{\cal O}(4) \right]  \,,  \quad \mbox{whereas}\label{e:p2}\\
\< f^p (\bn_o)\>&=&   f_0^p\left[1 -2p\ep^2\left\<\ka_1\de_1 \right\> + \frac{p\ep^2}{2}\<\de_2 \>  +\frac{p(p-1)\ep^2}{2}\< \de_1^2\> +{\cal O}(4)\right]\, .\label{e:p1}
\eea
It follows that, choosing $p-1=-\<\de_2 \> /\< \de_1^2\>$, we can avoid second-order corrections to $\avd{ f^p}$, while choosing $p-1=\big(4\left\<\ka_1\de_1 \right\> - \<\de_2 \>\big)/ \< \de_1^2\>$ we can avoid second-order corrections to $\< f^p(\bn_o)\>$.
Hence there is an optimal power of a given variable which removes corrections to the mean at second order.

More generally, for an arbitrary function $F(f)$ we obtain
\bea\label{e:Fmean}
\<F(f (\bn))\>= \left\<\;\overline{ F(f)}\;\right\> &=& F(f_0) +\frac{\ep^2}{2}\Big[F''(f_0)f_0^2 \<\de_1^2\> + F'(f_0)f_0\<\de_2\>\Big]
+{\cal O}(4) \,,  \quad \mbox{whereas} \\
\label{e:Fens}
\<F(f (\bn_o))\> &=& F(f_0)  +\frac{\ep^2}{2}\Big[F''(f_0)f_0^2\<\de_1^2\> + F'(f_0)f_0\big\{\<\de_2\>-4\<\ka_1\de_1\>  \big\}\Big]
+{\cal O}(4) \,.\eea 
Hence choosing $F''(f_0)f_0/F'(f_0)=-\<\de_2\>/\<\de_1^2\>$ avoids contributions from second-order perturbations to the mean over directions, $ \left\<\;\overline{ F(f)}\;\right\>$,  while choosing 
$F''(f_0)f_0/F'(f_0)=\big(4\left\<\ka_1\de_1 \right\> - \<\de_2 \>\big)/ \< \de_1^2\>$ avoids second-order contributions to $\<F(f (\bn_o))\>$.

\section{Application to specific observables}\label{s:ex}

We consider various examples and show explicitly how the two averaging procedures lead to different contributions from second-order perturbations.

\subsection{Distances, Angular Sizes and Standard Areas} \label{sec:distance}

The angular diameter distance up to second-order has been fully calculated in~\cite{chris1,BenDayan:2012wi,chris,giov}. Here we are interested in the dominant second-order terms: we want to know what is the maximum impact that second-order contributions can have on the mean distance. As discussed in~\cite{cmb2nd}, this means that we need to take into account only the terms with the maximal number of transverse derivatives. We can neglect contributions proportional to the gravitational potential and its time and radial derivatives, relative to transverse derivative of the gravitational potential:
\be
\Psi,\, \partial_t\Psi,\, \partial_r\Psi \ll  \partial_a\Psi\, ,
\ee
where $\partial_a\Psi\equiv \bfe_a ^i\partial_i\Psi$ and $\bfe_a =(\bfe_\theta, \bfe_\varphi)$ are orthogonal to the photon propagation $\bn_o$.

We calculate the angular diameter distance up to second order with these simplifications. Here we just present the result, the detailed calculation is given in~\cite{cmb2nd}. The angular diameter distance is given by the determinant of the 2x2 magnification matrix $\DD_{ab}$, defined as\footnote{Since photon propagation is conformally invariant, $\DD$ can be calculated in a non-expanding universe. For an expanding universe, we  multiply $\DD$  by the scale factor $a(\eta)$ (see appendix A of~\cite{shear}), which we rewrite as $1/(1+z)$ since perturbations in the redshift are negligible relative to the terms with four transverse derivatives.}
\be
\label{e:Dab}
\DD_{ab}= \bnabla_a\bal_b= \frac{\lambda}{1+z}\left(\begin{array}{cc}1-\kappa-\gamma^{(1)}&-\gamma^{(2)}-\omega\\ -\gamma^{(2)}+\omega& 1-\kappa+\gamma^{(1)} \end{array} \right)\,,
\ee
where $\omega$ is a curl component which vanishes at first order, $-2\ka =\bnabla\cdot\bal$ is the convergence and $\ga^{(1)}$ and $\ga^{(2)}$ are the shear components (from now on, we absorb the smallness parameter $\ep$ in the perturbation variables).
The magnification matrix obeys a second-order differential equation, which can be solved order by order in perturbation theory. The angular diameter distance is given by  
\be
\label{e:dA2}
d_A^2\simeq\frac{\chi^2}{(1+z)^2}\Big[1-2\kappa+\kappa^2-|\gamma|^2 \Big]\, ,
\ee
where $|\gamma|^2\equiv \big(\ga^{(1)}\big)^2+\big(\ga^{(2)}\big)^2$ and we have neglected $\omega^2$ which is of order 4 in perturbation theory.
We have also set $\lambda= \chi=\eta_0-\eta$, i.e. we neglect the perturbations of the affine parameter which contain less transverse derivatives than the deflection angle; $\eta_0$ and $\eta$ denote respectively the conformal time today and at the source position. Expanding $\ka$ to second order, $\ka=\ka_1+\ka_2/2$, and taking the square root of (\ref{e:dA2}),
\be\label{e:dAgen}
d_A = d_0\left[1 + \de_1 + \frac{1}{2}\de_2 +\OO(3)\right]= d_0\left[1-\kappa_1 -\frac{1}{2}\kappa_1^2 -\frac{1}{2}\ka_2+\frac{1}{2}(\kappa_1^2-|\gamma_1|^2) +\OO(3)\right]\, ,
\ee
where $d_0=\chi/(1+z)$ is the background distance (recall that we neglect perturbations in the redshift, which contain less transverse derivatives and are therefore subdominant).
As  $|\ga|$ enters squared, we need it only to first order, $|\ga|=|\ga_1| + \OO(2)$.

We are interested in the average of $d_A$. As demonstrated in~\cite{cmb2nd}, the second-order convergence $\kappa_2$ can be written as a total divergence, so it does not contribute: both the ensemble average and the directional average of $\kappa_2$ vanish:
\be
\<\kappa_2 \>=\avd{\kappa_2}=0\, .
\ee
Furthermore, as we show in~\cite{cmb2nd}, the combination $\kappa_1^2-|\gamma_1|^2$ is also a total divergence, so that
\be
\big\langle\kappa_1^2-|\gamma_1|^2 \big\rangle=\avd{\kappa_1^2-|\gamma_1|^2}=0\, .
\ee
If we take the ensemble average of the directional average, we obtain therefore 
\be
\label{dAdir}
\avd{d_A}=\<d_A(\bn)\> =d_0\left(1  -\frac{1}{2}\left\<\ka_1^2\right\>\right)\,.
\ee
This shows  that lensing {\it decreases the directional average} of the distance with respect to the background value~$d_0$.

On the other hand, if we calculate the ensemble average of the distance, we have a remaining second-order term given by~\eqref{correction}. Combining this second-order term with the first-order square, we obtain
\be
\label{dAexp}
\<d_A(\bn_o)\>=d_0\left(1+\frac{3}{2}\left\<\ka_1^2\right\>\right)\, .
\ee
This is the result discussed in~\cite{cmb}. It shows that lensing {\it increases the ensemble average} of the distance with respect to the background value $d_0$. 

The difference between \eqref{dAdir} and~\eqref{dAexp} can be interpreted in the following way. If we consider many realisations of a random but fixed line of sight, the fact that there are structures between the source and the observer will, on average, increase the distance. This means that structures generate with higher probability a de-focusing on a random line of sight. However, if we
average the distance over directions, under-densities which lead to de-focusing are competing with over-densities which lead to focusing. Since lensing also changes the distribution of lines of sight within a given solid angle, the contribution from under-densities does not exactly cancel the contribution from over-densities. On average, the latter dominates, leading to a decrease of the distance.

Taking an arbitrary power of $d_A$ to second order and making use of (\ref{e:p2}) and (\ref{e:p1}), with $\de_1=-\ka_1$ and $\de_2=-\ka_1^2 \,\,+$  divergence, we obtain
\bea
\<d_A^p(\bn)\>= \left\<\;\overline{ d_A^p}\;\right\> &=& d_0^p\left[1  +\frac{p(p-2)}{2}\left\<\ka_1^2\right\>+{\cal O}(4) \right] \,, \quad  \mbox{whereas}\label{dAdirp} \\
\<d_A^p(\bn_o)\> &=& d_0^p\left[1   +\frac{p(p+2)}{2}\left\<\ka_1^2\right\> +{\cal O}(4) \right] \,.
\eea
This shows that the ensemble average of $d_A^p$ experiences no second-order corrections for $p=-2$, while the ensemble average of the mean over directions of $d_A^p$ experiences no second-order corrections for $p=+2$. These results are completely consistent with~\cite{kibble}, noting that $\mu=d_0^2/\det \DD_{ab}=(d_0/d_A)^2$. The background distance is then given by
\be
d_0=\frac{\chi}{1+z}=\left[ \frac{p+2}{4}\left\<\;\overline{ d_A^p}\;\right\> - \frac{p-2}{4}\left\<\;{ d_A^p(\bn_o)}\;\right\> \right]^{1/p}\, .
\ee

From these results we can also calculate the observed angular sizes of standard objects on the sky determined via
\begin{align}
&\text{\em solid~angle}& &\<\De\Omega(\bn_o)\> = \left\<d_A^{-2}(\bn_o)\right\>\De S=\De\Omega_0,\hspace*{6cm} \\ && \text{while} ~~~& \<\overline{\De\Omega}\> = \<{d_A^{-2}(\bn)}\>\De S=\De\Omega_0 \Big(1+4\<\kappa_1^2\>\Big)\, .\\
&\text{\em linear~angle}&&\<\De\theta(\bn_o)\> = \left\<d_A^{-1}(\bn_o)\right\>\De L=\De\theta_0\left(1-\frac{1}{2}\<\kappa_1^2\>\right), \\ && \text{while}~~~& \<\overline{\De\theta}\> =\<d_A^{-1}(\bn)\>\De L= \De\theta_0\left(1+\frac{3}{2}\<\kappa_1^2\>\right)\, .
\end{align}
Here $\De S$ and $\De L$ are the standard area and standard length, and we have used (\ref{e:fbar}), i.e.,  $\<\;\overline{f}\;\> = \< f(\bn)\>$.
Only one of these is invariant: the expectation value of the angular size of a `standard area'. Any standard \emph{ruler}, whether averaged over directions or not, will receive corrections.  

\begin{figure}[!t]
\centering
\includegraphics[width=0.5\textwidth]{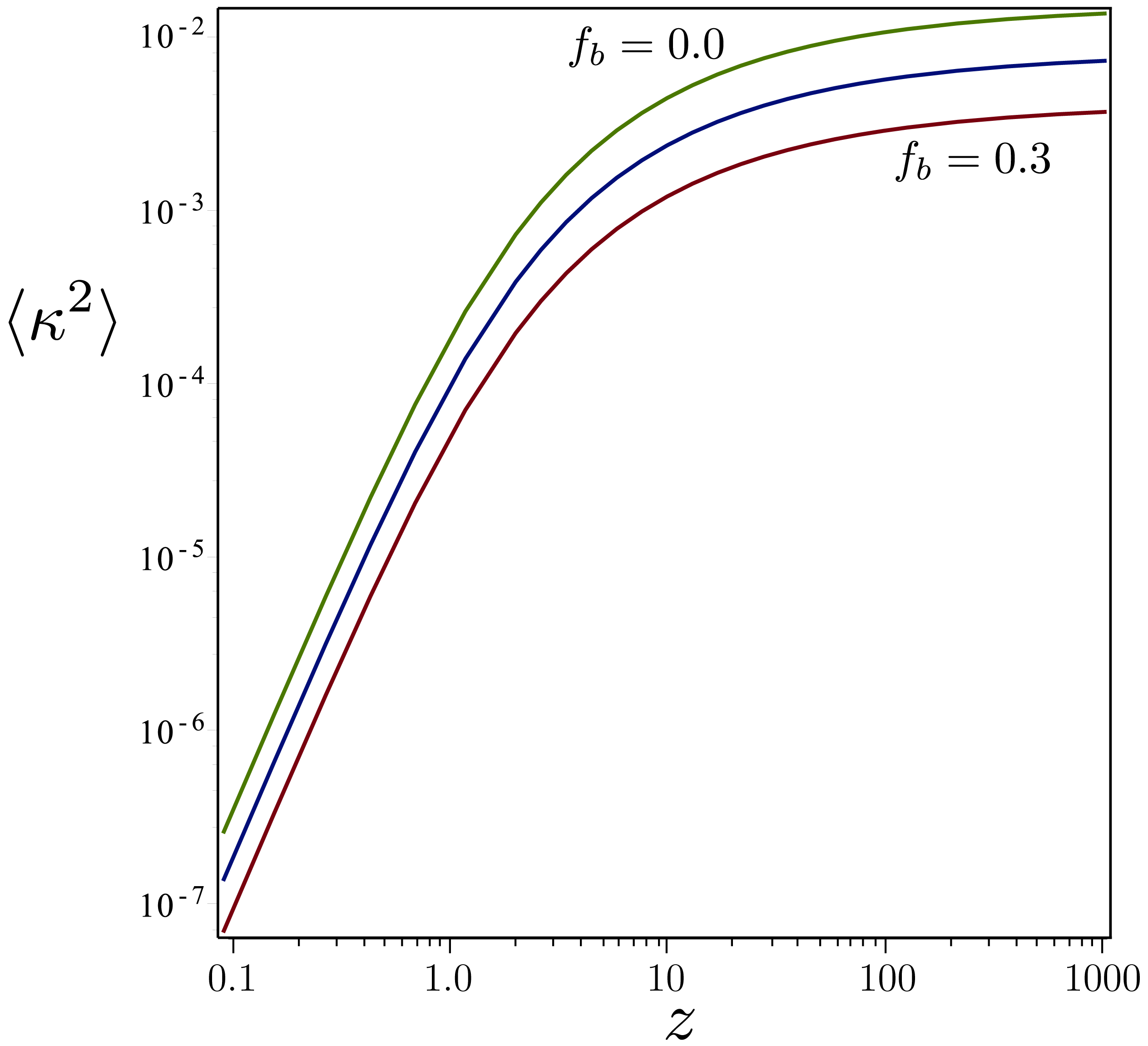}
\caption{\label{fig:kappasquare} The main contribution to $\langle \kappa_1^2\rangle$ comes from terms with 4 transverse derivatives, shown here for $\Omega_m=0.3, h=0.7$ and $n_s=0.96$. It is extremely sensitive to small-scale structure, illustrated here by considering different baryon fractions $f_b\equiv \Omega_b/\Omega_m=0,\,0.15,\, 0.3$. Higher $f_b$ damps small-scale power below 100\,Mpc through Silk damping, resulting in a smaller $\langle \kappa_1^2\rangle$. At $z\sim 1100$ an approximate dependence on model parameters is  $
 \langle \kappa_1^2\rangle\approx 0.42\, \Omega_m^{1.1f_b+1.8} h^{2.5f_b+3} n_s^5 e^{-12.6\Om_b h^2}$. Note that here we have used the linear power spectrum to calculate $\kap$. At high redshift, this is a good approximation, but at low redshift the linear expression underestimates the effect.}
\end{figure}

The observables discussed here are all affected by the square of the convergence $\kap$. The main contribution to $\kap$ can be calculated using the Limber approximation:
\bea
\kap&=&\left[\int_0^{\chi}d\chi'\frac{\chi-\chi'}{\chi\chi'}\Delta_\Omega\Psi(\chi)\right]^2\simeq
4\pi\sum_{\ell=0}^\infty \left[\frac{\ell(\ell+1)}{2\ell+1} \right]^2\int_0^\chi d\chi' \frac{(\chi-\chi')^2}{\chi'\chi^2}g^2(\chi')
\big(P_0 T^2\big)\left(k=\frac{\ell+1/2}{\chi'} \right)\, ,
\eea
where $\Delta_\Omega$ is the angular Laplacian, $T$ is the transfer function, $P_0$ is the power spectrum of the primordial gravitational potential and $g$ is the growth factor (see~\cite{cmb} for more detail). Figure~\ref{fig:kappasquare} shows $\kap$ as a function of redshift for different values of cosmological parameters. At the last scattering surface, $z\sim 1100$, $\kap$ reaches 0.6 percent. 

In the discussion above, we have considered only the perturbations with four transverse derivatives of the gravitational potential. A crucial consequence is that in this case, the second-order convergence $\kappa_2$ can be written as a total divergence, which vanishes on average. The only remaining contribution is therefore the square of the first-order convergence $\kap$. However, the full relativistic expression for the distance contains also second-order terms with two transverse derivatives of the gravitational potential and second-order terms with no transverse derivatives~\cite{chris1,BenDayan:2012wi,chris,giov}. 

The terms with no transverse derivatives are for example due to the integrated Sachs-Wolfe or the Shapiro time delay. These terms change the physical length of the photon geodesic between the source and the observer. They do not vanish on average and they affect therefore the mean distance to the source. Their amplitude is however of the order of the square of the gravitational potential, $\Psi^2\sim 10^{-10}$, i.e. much smaller than the first-order convergence square in \eqref{dAdir} and \eqref{dAexp}. In addition to the ISW and the Shapiro time delay, the distance receives also corrections from the Doppler terms. The square of the velocity is of the order $10^{-6}$ and the Doppler terms are therefore relevant only at very low redshift $z\lesssim 0.5$.

The terms with two transverse derivatives describe a coupling between the longitudinal and transverse deflections. For example, some of these terms are due to the fact that we average the distance over directions $\bn_o$ at a fixed value of the redshift. Since the redshift is itself perturbed, $z=z_0+\delta z$, we obtain contributions proportional to 
\be
d_A(z_0+\delta z,\bn_o)\sim d'_A(z_0,\bn_o)\cdot \delta z \sim d_0\, \kappa_1\, \frac{\delta z_1}{1+z}\, ,
\ee
i.e. contributions due to the fact that the distance is integrated on a perturbed surface, at a fixed redshift from the observer. As seen from Fig.~\ref{fig:kappasquare}, $\kap$ reaches $6\times 10^{-3}$ at very large redshift. The terms with two transverse derivatives are therefore roughly of the order $\kappa_1 \times \int \Psi\sim \sqrt{0.006}\times 10^{-5}\sim 8\times 10^{-7}$. It is therefore also justified to neglect these types of terms relative to the square of the convergence. 

Finally, let us mention that the distance contains also first-order contributions proportional to the gravitational potential at the observer, $\Psi_0$, and to the peculiar velocity at the observer, $\mathbf{v}_o\cdot\bn_o$. Whereas the velocity term vanishes under directional average, the potential term remains and contributes at the order of $10^{-5}$. This contribution is almost three orders of magnitude smaller than $\kap$ at high redshift. Below redshift 1 however, where $\kap$ is much smaller, this contribution may be relevant (see~\cite{tamara} for a detailed analysis of the impact of this term on supernovae measurements).

\subsection{Fluxes, Magnitudes and Standard Candles}

For a standard(-izable) candle such as a type Ia supernova, the relevant quantities are not the geometrical variables angle and area distance, but rather the flux of photons at the observer and the luminosity distance. Observers typically plot the distance modulus $\mu_m=m-M$ to a supernova at redshift $z$ of observed magnitude $m$ and true magnitude $M$, inferred via the observed flux $\FF$  and defined as follows\footnote{The distance modulus, denoted here by $\mu_m$, should not be confused with the lensing magnification $\mu$.},
\bea
\FF(z) &=& \frac{L}{4\pi d_L^2(z)} \,,\label{e:F} \\
\mu_m(z)-25 &=& 5\log\left[\frac{d_L(z)}{25\,\text{Mpc}}\right]  \,.  \label{e:mu}
\eea
Here $z$ is the redshift of the supernova (we neglect its perturbation which is justified only for $z\gtrsim 0.5$, see~\cite{BenDayan:2012wi,Ben-Dayan:2014swa,giov}), and $L$ is its intrinsic luminosity. Observers usually do not have many supernovae with the same redshift in different directions. Therefore, they directly fit the observed curve $\mu_m(z)$ to the corresponding curve for some background cosmology, without taking into account perturbations. First-order perturbations to the distance have been discussed and taken into account as a systematic error~\cite{Bonvin:2005ps,sne, hui}.
Here we discuss a shift of the mean value due to second-order fluctuations. We concentrate on redshifts $z>0.5$. The effect on the Hubble constant from close-by supernovae with $z<0.1$ is discussed in~\cite{Ben-Dayan:2014swa}.

The luminosity distance is related to the angular diameter distance through $d_L(z) = (1+z)^2d_A(z)$. Neglecting the perturbations in the redshift we have
\be
\label{eq:delta_dL}
\frac{\de d_L}{d_L}  = \frac{\de d_A}{d_A} = -\ka_1-\frac{1}{2}\ka_2  -\frac{1}{2}|\ga_1|^2 \, .
\ee
Inserting this into~(\ref{e:F}) and (\ref{e:mu}) and expanding up to second order we obtain
\bea
\FF(z)&=&\FF_0\Big(1+2\kappa_1+\kappa_2+3\kappa_1^2+|\gamma_1|^2 \Big)\, ,\label{eq:delta_F}\\
\mu_m(z)&=&\mu_{m0}(z)-5\kappa_1-\frac{5}{2}\Big(\kappa_2+\kappa_1^2+ |\gamma_1|^2\Big)\, .
\eea

The impact of second-order lensing on supernovae measurements then really depends on how observations are performed. 
If the observed flux of each supernova is directly compared with the intrinsic luminosity to extract the luminosity distance, then the relevant quantity is the expectation value of the flux. This is unaffected by second-order lensing, since $\<\ka_2\>=0$ and $\<\ka_1^2-|\ga_1|^2\>=0$:
\be
\<\FF(\bn_o)\>=\FF_0\,.
\ee 
On the other hand, if the distance modulus $\mu_m$ is used directly to extract the luminosity distance, then second-order lensing will systematically increase the expectation value
\be
\<\mu_m(z,\bn_o)\>= \mu_{m0}(z) +5\left\<\ka_1^2\right\>\, ,
\ee
leading to an overestimate of the distance $d_L$. Considering Fig.~\ref{fig:kappasquare}, we find that for $z<2$ the shift in $\mu_m$ is less than 0.003 and therefore will produce a shift in the dark energy equation of state around the percent level. However, since it is a shift with a definite sign leading to a slight overestimation of $\mu_m$ and hence of supernovae distances, this can bias parameter estimation if not taken into account.

The discussion above assumes that the luminosity distance is extracted individually from each supernova. If the sample is large enough, one can instead split the supernovae into $N$ bins of redshift and average the observed flux over all supernovae in the same bin.  This average can be considered as an approximate angular average, or equivalently expectation over true source positions (per redshift bin) for fixed size angular patches:
\be
\label{eq:flux}
\frac{1}{N}\sum_{i=1}^N\FF_i(z,\bn_o^i)=\frac{1}{4\pi}\sum_{i=1}^N\FF_i(z,\bn_o^i)\Delta\Omega_{\bn_o} \simeq  \avd{\FF(z)}=\<\FF(z,\bn)\>=\FF_0(z)\Big(1+4\<\kappa_1^2\>\Big)\,,
\ee
where we used \eqref{e:fcomp}.
We see that in this case the observed flux is increased by second-order lensing, leading to an underestimation of the luminosity distance. Hence even though averaging over supernovae in the same redshift bin has the advantage of decreasing the statistically uncertainty in the measurement of the flux, it has the disadvantage of introducing a systematic bias in the luminosity distance. This bias is also present if we average the distance modulus
\bea
\avd{ \mu_m(z)}  &=& \mu_{m0}(z) -5\left\<\ka_1^2\right\> \,,
\eea
and the distance is also underestimated in this case. From \eqref{dAdirp} we see that to fully exploit the potential of averaging over bins of redshift, without introducing any additional bias, we need to first extract the square of the luminosity distance for each supernova in the bin, and then take an average
\be
\avd{d^2_L}=d^2_{L0}\, .
\ee 

Finally, let us note that in practice, since supernovae are not perfect standard candles, their intrinsic luminosity is calibrated on a training subset, for which both the flux and the distance are known. From this subset a relation between the intrinsic (peak) luminosity and the shape of the light-curve (for example its width) is derived~\cite{calibration}. This relation is then used to determine the intrinsic luminosity of the other supernovae. One can then wonder whether this relation is affected by second-order lensing, leading to a systematic bias in the determination of the intrinsic luminosity and consequently to an error in the distance measurements. This is fortunately not the case. Indeed, in the training subset, both the distance and the flux are affected in a consistent way by lensing\footnote{The distance to supernovae in the training subset is indeed usually measured through the host galaxy using its surface brightness fluctuation or the Tully-Fisher relation to determine the galaxy intrinsic luminosity. Photons emitted by the galaxy follow the same path as the supernova's photons and experience therefore the same lensing corrections.} (following \eqref{eq:delta_dL} and \eqref{eq:delta_F}), leaving the intrinsic luminosity unchanged
\be
L=4\pi d_L^2 \FF=4\pi d_{L0}^2\FF_0=L_0\, .
\ee
The shape of the light-curve is also unaffected, since lensing is not expected to vary during the time-scale of the supernova explosion. As a consequence, lensing can induce a constant shift in the amplitude of the light-curve, but it cannot generate a change in the shape. Therefore the relation between the shape of the light-curve and the intrinsic luminosity inferred from the training subset is not biased by second-order lensing.

\subsection{CMB angular power spectrum} 

As a last example, we discuss what happens with the CMB. In~\cite{cmb2nd}, we  calculated the contribution of the shear and the convergence to the angular power spectrum of the CMB, up to second order in perturbation theory. We found that the lensed power spectrum $\tilde D(\ell)\equiv \ell^2 \tilde C(\ell)$, calculated for a constant magnification matrix, is related to the unlensed power spectrum $D(\ell)$ through
\be
\label{Dl}
\tilde D(\ell)=D(\ell)+\left[\kappa+\kappa^2+\frac{1}{4}|\gamma|^2 \right] \ell D'(\ell)
+\frac{1}{2}\left[\kappa^2+\frac{1}{2}|\gamma|^2 \right] \ell^2 D''(\ell)\, .
\ee
If we take a directional average of \eqref{Dl}, as we do in~\cite{cmb2nd}, i.e. we average the shear and convergence over different parts of the sky, we obtain
\be
\label{Dlavfin}
\avd{\tilde D(\ell)}=D(\ell)+\frac{5}{4}\kap \ell D'(\ell)
+\frac{3}{4}\kap \ell^2 D''(\ell)\, .
\ee
If instead we take an ensemble average, i.e. we average over all possible realisations of the shear and convergence fields, we obtain
\be
\label{Dlavexp}
\left\<\tilde D(\ell) \right\>=D(\ell)-\frac{3}{4}\kap \ell D'(\ell)+\frac{3}{4}\kap \ell^2 D''(\ell)\, .
\ee
The smoothing term $\ell^2 D''(\ell)$ is the same in the two cases: it shows that lensing decreases the amplitude of the peaks. The pure displacement term $\ell D'(\ell)$ on the other hand is different: it is positive if we take a directional average but negative if we take an ensemble average. This is completely equivalent to the effect on the distance described in section~\ref{sec:distance}: the ensemble average of the distance is increased whereas the directional average is decreased. Since the CMB is averaged over directions, the correct averaging procedure is given by \eqref{Dlavfin}, leading to a shift of the peaks to lower multipoles\footnote{Note that the shift in the observed position of the peaks is governed both by the displacement term and by the smoothing term. Since the smoothing term dominates over the displacement term, the shift is also to lower multipoles in \eqref{Dlavexp}, but its amplitude is smaller.}. As discussed in detail in~\cite{cmb2nd}, this shift is consistently included in standard Boltzmann codes (such as CAMB~\cite{camb} or CLASS~\cite{class}), since it is due to the square of the first-order convergence. The only term in \eqref{Dl} neglected by current CMB analyses is the contribution from the second-order convergence $\kappa_2$. However, as already mentioned before, the dominant terms in $\kappa_2$ (those with 4 transverse derivatives) vanish on average, and the subdominant terms are negligible, roughly $10^4$ times smaller than the first-order convergence square (see discussion at the end of section~\ref{sec:distance}).

Finally, let us mention that the standard way of calculating the CMB angular power spectrum is not through~\eqref{Dlavfin} but rather through a calculation of the lensed multipoles $a_{\ell m}$. This calculation automatically selects the correct averaging procedure since the $a_{\ell m}$ are defined through a (weighted) average over directions
\be
a_{\ell m}=\int d\Omega_{\no}Y_{\ell m}(\no)T(\no)\, .
\ee
The angular power spectrum is given by
\be
\<a_{\ell m}a_{\ell' m'} \>=C_\ell \delta_{\ell \ell'}\delta_{mm'}\, .
\ee
The ensemble average is therefore automatically taken {\it after} the integral over directions.

\section{Conclusions}\label{s:con}

In this paper we have shown that at second order in perturbation theory, averaging over the observed directions and taking an ensemble average are two distinct procedures, which do not commute. This comes from the fact that the observed direction is lensed -- hence it is itself  perturbed -- and its perturbation may well be correlated with the observable under consideration. 

This is especially relevant for distance measurements, the perturbations of which are intimately related to lensing. In particular, we have shown that the distance to the last scattering surface is decreased by lensing if one takes an average over directions, whereas it is increased if one takes an ensemble average. In a companion paper~\cite{cmb2nd} we argue that the directional average is relevant for CMB observations and that consequently second-order lensing shifts the position of the peaks to lower multipoles. However, as we showed there, the change in the distance captures the essence of the change to the peaks but does not accurately capture the shift or damping of the peaks. Consequently, the CMB sound horizon does not act as a standard ruler when lensing is present. 

For the analysis of diffuse observables like the CMB (actually also the BAO's), correlation functions are calculated and fitted to the model. Moreover, in multipole space, the average over directions is always taken before the ensemble average, and no ambiguity arises concerning the averaging procedure. For standard rulers and candles which are point sources, it is much more subtle to remove the bias arising from lensing in going from the observable to the model, and the distance measure chosen must be carefully understood. In particular, the notion of standard ruler should be extended to a standard area as the expectation value of an observed solid angle is preserved under lensing, which is not the case for an observed linear angle.

We have also shown that for specific observables there exist functions which do not acquire corrections from second-order perturbations. It is therefore a good observational strategy to consider these functions of the observables in order to avoid systematic errors from second-order perturbations.

~\\\\{\it Note added:} While this paper was being finalised, an independent paper on the same topic appeared~\cite{kaiser}.

~\\{\bf Acknowledgements:} We thank Giovanni Marozzi for interesting and helpful discussions. RD is supported by the Swiss National Science Foundation.
CC, RM and OU are supported by the South African National Research Foundation.
RM and OU  are supported by the South African Square Kilometre Array Project and  RM acknowledges support from the UK Science \& Technology Facilities Council (grant ST/K0090X/1).


\end{document}